\documentstyle[12pt]{article}

\newcommand{\sect}[1]{\setcounter{equation}{0}\section{#1}}

\def\beq{\begin{equation}}
\def\eeq{\end{equation}}
\def\bea{\begin{eqnarray}}
\def\eea{\end{eqnarray}}
\def\bq{\begin{quote}}
\def\eq{\end{quote}}

\newcommand{\EQ}{\begin{equation}}
\newcommand{\EN}{\end{equation}}
\newcommand{\ena}{\end{eqnarray}}

\renewcommand{\a}{\alpha}
\renewcommand{\b}{\beta}

\renewcommand{\d}{\delta}
\newcommand{\th}{\theta}

\newcommand{\pa}{\partial}
\newcommand{\g}{\gamma}

\newcommand{\e}{\epsilon}

\newcommand{\m}{\mu}

\renewcommand{\o}{\omega}

\newcommand{\del}{\partial}

\newcommand{\NP}[1]{Nucl.\ Phys.\ {\bf #1}}

\renewcommand{\thefootnote}{\fnsymbol{footnote}}

\begin{document}
\newpage
\begin{titlepage}
\begin{flushright}
{CERN-TH.6336/91}\\
{IFUM 412/FT}\\
{hepth@xxx/9202069}
\end{flushright}
\vspace{2cm}
\begin{center}
{\bf {\large QUANTUM CONSERVED CURRENTS IN}} \\
\vspace{.1in}
{\bf {\large AFFINE TODA THEORIES }} \\
\vspace{1.5cm}
{G.W. DELIUS and
M.T. GRISARU}\footnote{On leave from Brandeis University, Waltham, MA 02254,
USA\\Work partially supported by the National Science Foundation under
grant PHY-88-18853 and by INFN}\\
\vspace{1mm}
{\em Theory Division, CERN, 1211 Geneva 23, Switzerland}\\

\vspace{5mm}
and

\vspace{5mm}
{D. ZANON} \\
\vspace{1mm}{\em Dipartimento di Fisica dell' Universit\`{a} di Milano and} \\
{\em INFN, Sezione di Milano, I-20133 Milano,
Italy}\\
\vspace{1.1cm}
{\bf{ABSTRACT}}
\end{center}
\bq
We study
 the renormalization and conservation at the quantum level of
higher-spin currents in affine Toda theories
with particular emphasis on the nonsimply-laced cases.
For specific examples, namely the spin-3 current
for the
$a_3^{(2)}$ and $c_2^{(1)}$ theories, we prove conservation to all-loop order,
thus establishing the existence of
factorized S-matrices. For these theories, as well as the simply-laced
$a_2^{(1)}$ theory, we compute one-loop corrections to the corresponding
higher-spin charges and study charge conservation for the three-particle
vertex function.
For the $a_3^{(2)}$ theory
we show that although the current is conserved,
anomalous threshold
singularities spoil the conservation of the
corresponding charge for the on-shell vertex function, implying a breakdown
of some of the bootstrap procedures commonly used in determining the
exact S-matrix.
\eq

\vfill

\begin{flushleft}
CERN-TH.6336/91\\
IFUM 412/TH\\
December 1991
\end{flushleft}
\end{titlepage}

\renewcommand{\thefootnote}{\arabic{footnote}}
\setcounter{footnote}{0}
\newpage

\sect{Introduction}

Affine Toda field theories describe a class of massive systems which are
classically integrable; from the identification of the field equations
as compatibility conditions for a Lax pair follows the existence of
an infinite number of classically conserved higher-spin currents
$J_+^{(s)}$, $J_-^{(s)}$ \cite{1}
\EQ
\pa_-J_+^{(s)}+\pa_+J_-^{(s)}=0
\EN
Unless the conservation is spoiled by quantum anomalies,
the existence of the corresponding
charges $Q^{(s)}$ implies the factorization and
elasticity of the S-matrix \cite{2,3}. Unitarity and a bootstrap
principle lead then to its exact determination.

The quantum-conservation of the higher-spin currents has been
investigated by a number of authors for the simplest of the affine Toda
theories, namely the sine- (or sinh-) Gordon system, as well as its
supersymmetric counterpart \cite{4,5}. By using BPHZ techniques, or
conformal field theory methods, it has been shown that renormalized
currents can be defined which indeed are conserved.
Similar
investigations have been carried out in the context of perturbed
conformal field theories \cite{6}.

Our interest in the quantum-conservation arose out of work
on exact S-matrices for
nonsimply-laced affine Toda theories. For simply-laced affine Toda
theories exact S-matrices were constructed some time ago \cite{7,7.5}
but similar constructions had failed for the nonsimply-laced cases, suggesting
that perhaps the classical integrability breaks down at the quantum level.
It was the realization that
higher-spin quantum-conserved currents exist that gave the
impetus for a new, successful effort, to determine hitherto unknown
S-matrices for these systems \cite{8}. In this paper we present details of the
construction of such currents and of corresponding charges. We also
discuss the issue of charge conservation in the presence of anomalous
threshold singularities, and implications for the bootstrap program
of the S-matrix.

Because explicit formulas are not known for currents of
high spin and also to avoid algebraic complexity, we have restricted
ourselves to the lowest nontrivial spin $s \geq 2$ (spin $s=1$
corresponds to the stress tensor), for nonsimply-laced
Toda theories with only two bosonic
fields, namely the spin-3 current for the
$a_3^{(2)}$ and $c_2^{(1)}$ theories. However, to set the notation and
for comparison, we also consider the spin-2 current for the $a_2^{(1)}$
simply-laced theory (the results are easily extended
to the simply-laced
$a_n^{(1)}$ case for any $n$). In general, two-dimensional power
counting arguments and Lorentz
invariance suggest that any violation at the quantum level
of the classical conservation laws
for a spin $s$ current can occur at most at $l=s-1$ loops, and from
a restricted class of diagrams; for bosonic theories they are diagrams
obtained by Wick-contracting the currents with just one factor of the
interaction lagrangian. We are thus able to obtain
all-loop results for the absence of anomalies in the currents.

The affine Toda theories have interaction lagrangians which are exponentials
of sums of fields. From these one can separate the quadratic parts which
correspond to mass terms.
We have studied the conservation of currents by using two different methods:
massive perturbation theory and BPHZ
techniques, and massless perturbation theory which  treats the whole
exponentials as interaction terms and is
essentially equivalent to OPE techniques. Both
methods lead to the same conclusions, but calculations using
massless perturbation methods
are much simpler beyond one loop and are the ones that will be presented
here.

Our paper is organized as follows: in Section 2 we set up the notation
and describe the general procedure for constructing quantum-conserved
currents. In Section 3 we carry out the construction for the cases
mentioned above; the {\em exact} currents we obtain consist of classical
parts, plus quantum corrections. In Section 4 we define, and compute
up to one-loop order the corresponding charges. In Section
5 we show that the one-loop corrections to the classical charges
are consistent with quantum corrections to the masses of the
particles, by studying the charge-conservation relations for the three-point
on-shell vertex functions. In Section 6 we show how these relations are
affected by the presence of anomalous threshold singularities, and discuss
implications for the S-matrix bootstrap principle. Section 7 contains
discussion
and
conclusions.

\sect{Massless perturbation theory conventions}

We consider lagrangians of the form
\EQ
\b^2 {\cal L} = -\frac{1}{2} \vec{\phi}\Box \vec{\phi} - \m^2 \sum_i
q_ie^{\vec{\a}_i \cdot \vec{\phi}}
\EN
Here the $\vec{\a}_i$ are the positive simple roots of a rank $n$ Lie
algebra augmented by (the negative of) a maximal root and $\vec{\phi}
=(\phi_1,\phi_2, \dots \phi_n)$ are bosonic fields describing $n$
massive particles. The quantum lagrangian is defined by normal-ordering
the exponentials after which all Green's functions of the basic fields
are free of any divergences. The Ka$\check{c}$ labels $q_i$ are
chosen so that the one-point functions vanish.
$\b$ is the coupling constant, and $\m$ sets
the mass scale; we choose $\m =1$. At the classical level these theories
possess conserved currents of spins $s$ equal to the exponents of the
algebra modulo the Coxeter number \cite{1}.

We work in
Minkowski space and use light-cone coordinates; for
convenience, whenever possible, we use a notation reminiscent of
Euclidean space
\bea
z \equiv x^+ &=& \frac{1}{\sqrt{2}}(x^0+x^1) ~~~~,~~~~\bar{z}
\equiv x^- =\frac{1}{\sqrt{2}}
(x^0-x^1) \nonumber\\
\pa \equiv \pa_+ &=& \frac{1}{\sqrt{2}}(\pa_0+\pa_1)
{}~~~~,~~~~\bar{\pa} \equiv \pa_-=\frac{1}
{\sqrt{2}}(\pa_0-\pa_1) \nonumber\\
\Box &=& 2\pa \bar{\pa}
\ena
and also
\EQ
J^{(s)} \equiv J^{(s)}_+ = \frac{1}{\sqrt{2}}(J^{(s)}_0+J^{(s)}_1)
{}~~~~,~~~~\bar{J}^{(s)} \equiv J^{(s)}_-=\frac{1}
{\sqrt{2}}(J^{(s)}_0-J^{(s)}_1)
\EN
We perform calculations in $x$-space with massless bosonic propagators
\EQ
\left\langle \phi (z,\bar{z}) \phi (0,0)\right\rangle  =
-\frac{\b^2}{4\pi}\log (2z \bar{z})
\EN

We consider currents of the form
\bea
J^{(s)}&=& \sum\,a_{pq}\,\del^{p_1}\phi_{q_1}\cdots
\del^{p_n}\phi_{q_n}\nonumber\\
\bar{J}^{(s)}&=& \sum_i\left(\sum\,b_{itr}\,\del^{t_1}\phi_{r_1}\cdots
\del^{t_n}\phi_{r_n}\right)
\,e^{\vec{\a}_i\cdot\vec{\phi}}
\eea
with $\sum p_i =s+1$, $\sum t_i =s-1$ and
with coefficients $a_{pq}= a_{pq}^{(0)}+\b^2 a_{pq}^{(1)} +\cdots$, etc..  The
coefficients $a_{pq}^{(0)}$ and $b_{itr}^{(0)}$ are determined (up to the
shift $J\rightarrow J+\del\Lambda$, $\bar{J}\rightarrow \bar{J}
-\bar{\del}\Lambda$) by the classical conservation laws
$\bar{\pa}J^{(s)}+\pa \bar{J}^{(s)}=0$.  At the quantum level we compute
\EQ
\bar{\pa}_z \left\langle J(z,\bar{z})\right\rangle  \equiv
\bar{\pa}_z \left\langle J(z,\bar{z})\
\exp \left(i \int d^2w {\cal L}_{int} (w,\bar{w})\right) \right\rangle _0
\EN
Potential anomalies would correspond to {\em local} contributions
which cannot be written as
the $\pa$-derivative of some suitable expression. We attempt to determine
the coefficients $a_{pq}$ to cancel such contributions.

For the lagrangian in eq. (2.1) we need only consider
expanding the exponential in eq. (2.6) to
first order in ${\cal L}_{int}$. Since the component
$J^{(s)}$ of the  currents contains exclusively terms
of the form $\pa^p \phi$ the Wick contractions in the above matrix
element will lead to a sum of terms of the form
\EQ
\int d^2w A(z, \bar{z})
\frac{1}{(z-w)^n} B(w,\bar{w})
\EN
where $A$, $B$, are products of fields and their  $\pa$ derivatives.

Since any potential anomalies are local, the $\bar{\pa}$-derivative
in eq. (2.6)
must act on the $(z-w)^{-n}$ factor. We use the standard prescription
\EQ
\bar{\pa}_z \frac{1}{(z-w)^n} =\frac{2\pi i}{(n-1)!} \pa_w^{n-1}
\d^{(2)}(z-w)
\EN
so that we are led to consider a sum of terms of the form
\EQ
A(z, \bar{z})\pa^{n-1}B(z,\bar{z})
\EN
that we should be able to rewrite
as some $\pa \bar{J}(z,\bar{z})$. We are not interested in the
actual form of $\bar{J}$, and so in our calculations we shall freely
"integrate by parts" on $z$, i.e. drop total $\pa$ derivatives.

A  spin $s$ current contains at most $s+1$ factors of fields. Therefore
the Wick contractions lead to contributions only up to $s$ loops.
However, the $s$-loop contribution is automatically a total
derivative so need not be considered further.
Thus, for example,
\nobreak{
\bea
&&\bar {\pa}_z \left\langle  (\pa \phi)^4 \left( -\frac{i}{\b^2}
\int d^2w e^{\phi} \right) \right\rangle
 \nonumber\\
&&~~~~=-i\bar{\pa }_z \int d^2w \left[4(\pa \phi)^3
 \frac{-1}{4\pi (z-w)} +6(\pa \phi )^2 \frac{\b^2}{16\pi^2 (z-w)^2} \right.
\nonumber\\
&&\left.~~~~~~~~~~~~~~
 +4 \pa \phi \frac{-\b^4}{64\pi^3(z-w)^3} +\frac{\b^6}{256\pi^4
(z-w)^4}\right]
e^{\phi} \nonumber\\
&&~~~~ \leadsto -2 (\pa \phi)^3 e^{\phi} -\frac{3\b^2}{4\pi}(\pa \phi)^2 \pa
 e^{\phi}
-\frac{\b^4}{16\pi^2}\pa \phi \pa^2 e^{\phi}
-\frac{\b^6}{768\pi^3}\pa^3e^{\phi}
\nonumber\\
&&~~~~ \leadsto [-4-\frac{3\b^2}{2\pi}
+\frac{\b^4}{16\pi^2}]\pa^3 \phi e^{\phi}
\ena}
where we have dropped the $\b^6$ contribution which is manifestly a
total derivative, and also used the identities (which are valid up to
total $\pa$ derivatives)
\bea
\pa (\pa^2 \phi e^{\phi} ) &=&( \pa^3\phi +\pa^2\phi \pa \phi )e^{\phi}
\sim 0 \nonumber\\
\pa ((\pa \phi )^2 e^{\phi}) &=& (2\pa \phi \pa^2\phi +(\pa \phi )^3)e^{\phi}
\sim 0
\ena

\sect{Quantum-conserved currents}

In this section we construct quantum-conserved currents for the $a_2^{(1)}$
simply-laced affine Toda theory, and for the nonsimply-laced $a_3^{(2)}$
and $c_2^{(1)}$ theories. These are theories with two bosonic fields,
general enough to illustrate the issues we wish to discuss. All these theories
have a conserved spin $s=1$ current, namely the stress-tensor $T\equiv T_{++}$,
$\bar{T} \equiv T_{+-}$, satisfying $\bar{\pa}T+\pa \bar{T}=0$.

\subsection{The $a_2^{(1)}$ affine Toda theory}

The lagrangian for this simply-laced theory is given by
\bea
\b^2{\cal L} &=& -\frac{1}{2}\phi_1 \Box \phi_1
-\frac{1}{2}\phi_2 \Box \phi_2 -e^{\sqrt{2}\phi_1}
\nonumber\\
&& -e^{-\frac{1}{\sqrt{2}}\phi_1 -\sqrt{\frac{3}{2}}\phi_2}
-e^{-\frac{1}{\sqrt{2}}\phi_1 +\sqrt{\frac{3}{2}}\phi_2}
\ena
General considerations \cite{1} establish the existence of a classically
conserved spin-2 current.

We consider, based on the $\phi_2 \rightarrow -\phi_2$ symmetry of the
lagrangian,
\EQ
J^{(2)}=A(\pa \phi_1)^2\pa \phi_2 +B (\pa \phi_2)^3 +C \pa \phi_1
\pa^2 \phi_2
\EN
Wick-contracting with the interaction lagrangian we obtain ($\a \equiv \frac
{\b^2}{2\pi}$)
\bea
&&\bar{\pa}\left\langle J^{(2)} \left(-\frac{i}{\b^2} \int
d^2w e^{\sqrt{2}\phi_1}\right)\right\rangle  \leadsto
\left[ (1+\frac{\a}{2})A -\frac{C}{\sqrt{2}}\right] \pa^2\phi_2
e^{\sqrt{2}\phi_1} \nonumber\\
&&\bar{\pa}\left\langle J^{(2)} \left( -\frac{i}{\b^2}
\int d^2w e^{-\frac{1}{\sqrt{2}}\phi_1
-\sqrt{\frac{3}{2}}
\phi_2} \right)\right\rangle  \nonumber\\
&&\leadsto \left[ \frac{\sqrt{3}}{2}
\left((1+\frac{\a}{2})A -\frac{C}{\sqrt{2}}\right)\pa^2 \phi_1
+\left(\frac{\a}{8}A +(\frac{3}{2}+\frac{9}{8}\a )B +
\frac{1}{2\sqrt{2}}C \right)\pa^2\phi_2 \right.\nonumber\\
&&~~~~~~~~~~~~~~~~~~~~~~~~-\left. \frac{1}{2\sqrt{2}}\left( A
+3B\right)\pa \phi_1 \pa \phi_2\right]
e^{-\frac{1}{\sqrt{2}}\phi_1 -\sqrt{\frac{3}{2}}
\phi_2}
\ena
where we have kept only the local part of the left-hand-side,
as in eq. (2.10).
We do not need to calculate separately contractions
with the third exponential in the lagrangian because of the symmetry under
$\phi_2 \rightarrow - \phi_2$.
We have made use of
identities (up to total $\pa$ derivatives), similar to the ones in eq. (2.11).

Current conservation requires that the terms on the right-hand-sides in
eq. (3.3), which are not total $\pa$ derivatives of some $\bar{J}^{(s)}$,
  vanish:
\bea
&& (1+\frac{\a}{2})A -\frac{1}{\sqrt{2}}C=0 \nonumber\\
&&\frac{\a}{8}A +(\frac{3}{2} +\frac{9}{8}\a
)B +\frac{1}{2\sqrt{2}}
C=0 \nonumber\\
&&A+3B=0
\ena
These three homogeneous equations with three unknowns
can be satisfied nontrivially since the determinant of the system vanishes
and thus we find, up to an overall normalization factor, a quantum-conserved
spin-2 current
\EQ
J^{(2)}= (\pa \phi_1)^2 \pa \phi_2 -\frac{1}{3}(\pa \phi_2)^3
+\sqrt{2}(1+\frac{\b^2}{4\pi}) \pa \phi_1 \pa^2\phi_2
\EN

The terms independent of $\b^2$ give of course the $J$ component of
the classical
conserved current.

We emphasize that it was not obvious a priori that the equations for
the coefficients $A,B,C$ would have a nontrivial solution so that a
quantum-conserved current exists.
However this result was known, and also expected since S-matrices
for the $a_n^{(1)}$
series had been constructed \cite{7}, implying the existence of
higher-spin quantum-conserved
charges.
In fact the result in eq. (3.5) can be generalized
to the $a_n^{(1)}$ theories
for any $n$: using the same techniques it is easy to
show that the current
\EQ
J^{(2)}= K_{abc}\pa \phi_a \pa \phi_b \pa \phi_c +K_{ab} \pa \phi_a
\pa^2\phi_b
\EN
is quantum-conserved provided the coefficients satisfy
\EQ
(1+\frac{\b^2}{4\pi })K_{abc}\a_a^i = \frac{1}{6}(K_{ab}\a_a^i \a_c^i
+K_{ac} \a_a^i\a_b^i )
\EN
where $\vec{\a}^i$ are the roots of the $a_n^{(1)}$ theory normalized to
$|\vec{\a}^i|^2 =2$. Since at the classical level a spin-2 conserved current is
known to exist \cite{1}, the above
equations have a solution if one drops the $\b^2$
term, with some coefficients $K^{(0)}_{abc}$, $K^{(0)}_{ab}$. It is
obvious then that the quantum current also exists, with
$K_{abc}=K^{(0)}_{abc}$, $K_{ab}= (1+\frac{\b^2}{4\pi}) K^{(0)}_{ab}$.
We note that the result in eq. (3.7), with the common factor
$1+\frac{\b^2}{4\pi}$ for any $i$, holds only
because $a_n^{(1)}$ is simply-laced.

\subsection{The $a_3^{(2)}$ affine Toda theory}

For this nonsimply-laced theory the lagrangian is
\EQ
\b^2{\cal L} =-\frac{1}{2}\phi_1 \Box \phi_1
-\frac{1}{2}\phi_2 \Box \phi_2 -e^{-\phi_1 -\phi_2}
-e^{-\phi_1+\phi_2} -e^{2\phi_1}
\EN
It has a spin-3 classically conserved current \cite{1}, so we start (using also
the symmetry of the lagrangian under $\phi_2 \rightarrow - \phi_2$) with
\EQ
J^{(3)}=A(\pa \phi_1)^2(\pa \phi_2)^2 +B(\pa \phi_1)^4 +C(\pa \phi_2)^4
+D \pa \phi_1 \pa \phi_2 \pa^2 \phi_2 +E(\pa^2\phi_1)^2 +F(\pa^2
\phi_2)^2
\EN

We find, writing the results in terms of an independent set (up to total
$\pa$ derivatives)
\bea
&&\bar{\pa}\left\langle J^{(3)}
\left(-\frac{i}{\b^2} \int d^2w e^{2\phi_1}\right)\right\rangle  \nonumber\\
&&~~~~ \leadsto \left(
\left[2(1+\a )A -D\right] \pa^2
\phi_2 \pa \phi_2
-\left[2(1+3 \a +\a^2)B -2E \right] \pa ^3\phi_1 \right)
e^{2\phi_1}
\nonumber\\
&&\bar{\pa}\left\langle J^{(3)}\left(-\frac{i}{\b^2}
\int d^2w e^{-\phi_1-\phi_2}\right) \right\rangle  \nonumber\\
&&~~~~\leadsto
\left( \left[ (1+\frac{\a}{2})A -(6+3\a )B
-\frac{1}{2}D \right]
\pa^2\phi_1 \pa \phi_2\right. \nonumber\\
&&~~~\left. +\left[(1+\frac{\a}{2})A -2B
 -(4+3\a )C -\frac{1}{2}D\right] \pa^2
\phi_2 \pa \phi_1 \right.\nonumber\\
&&~~~~\left.+\left[(\frac{\a}{2}
+\frac{\a^2}{8})A +(4+3\a +\frac{\a^2}{4})B
-\frac{\a}{8}D -E \right] \pa^3 \phi_1 \right.\nonumber\\
&&~~~~\left. +\left[(\frac{\a}{2}
+\frac{\a^2}{8})A +(4+3\a +\frac{\a^2}
{4})C +(\frac{1}{2}+\frac{\a}{8})D-F \right] \pa^3\phi_2 \right.
\nonumber\\
&&~~~~\left. +\left[ 2(B-C)\right](\pa \phi_2)^2 \pa \phi_1 \right)
e^{-\phi_1-\phi_2}
\ena
Again, the remaining exponential need not be considered because of the
$\phi_2 \rightarrow - \phi_2$ symmetry.

Requiring that the right hand side vanish we obtain
a set of equations for the coefficients $A,B,....F$, which have a
nontrivial solution leading to the quantum-conserved current
\bea
J^{(3)} &=& (1+\frac{\b^2}{4\pi})(\pa \phi_1)^2(\pa \phi_2)^2
-\frac{\b^2}{24\pi}(\pa \phi_1)^4 -\frac{\b^2}{24\pi}(\pa \phi_2)^4
\nonumber\\
&&+(2+\frac{3\b^2}{2\pi} +\frac{\b^4}{4\pi^2})\pa \phi_1 \pa \phi_2 \pa^2
\phi_2 - \frac{\b^2}{24\pi}(1+\frac{3\b^2}{2\pi}+\frac{\b^4}{4\pi^2})
(\pa^2\phi_1)^2 \nonumber\\
&&+(1+\frac{23\b^2}{24\pi}+\frac{\b^4}{4\pi^2}+\frac{\b^6}
{48\pi^3})(\pa^2\phi_2)^2
\ena
We note that in addition to the renormalization of terms already present
in the classical current new terms are generated at the quantum level. In
particular, the appearance of the $(\pa^2\phi_1)^2$ term has important
consequences that will be discussed later on.

The existence of the quantum-conserved current $J^{(3)}$ implies that
this theory must have factorizable, elastic S-matrices.

\subsection{The $c_2^{(1)}$ affine Toda theory}

Again this is a nonsimply-laced theory, with the lagrangian given by
\EQ
\b^2{\cal L}= -\frac{1}{2}\phi_1 \Box \phi_1
-\frac{1}{2}\phi_2 \Box \phi_2-e^{\sqrt{2}(\phi_1
-\phi_2)} -e^{-\sqrt{2}(\phi_1+\phi_2)} -2e^{\sqrt{2}\phi_2}
\EN
We consider the spin-3 current, even in $\phi_1$,
\EQ
J^{(3)}=A(\pa \phi_1)^4 +B(\pa \phi_2)^4 +C(\pa \phi_1)^2(\pa \phi_2)^2
+D(\pa \phi_1)^2\pa^2\phi_2 +E(\pa^2\phi_1)^2+F(\pa^2\phi_2)^2
\EN
We obtain
\bea
&&\bar{\pa}\left\langle J^{(3)} \left(-\frac{2i}{\b^2} \int d^2w
e^{\sqrt{2}\phi_2}\right)\right\rangle  \nonumber\\
&&~~~~\leadsto
\left(\left[-\sqrt{2}(4+6\a +\a^2)B
+2\sqrt{2}F\right]\pa^3\phi_2
+\left[(4+2\a )C
+2\sqrt{2}D\right]\pa^2\phi_1\pa \phi_1
\right)e^{\sqrt{2}\phi_2} \nonumber\\
&&\bar{\pa}\left\langle J^{(3)} \left(-\frac{i}{\b^2}\int d^2w
e^{\sqrt{2}(\phi_1-\phi_2}\right)\right\rangle  \nonumber\\
&&~~~~ \leadsto
\left(\left[(6+6\a )A-(1+\a )C
-\sqrt{2}D\right]\pa^2\phi_1 \pa \phi_1 \right. \nonumber\\
&&~~~~\left. +\left[2A+(4+6\a )B-(1+\a )C
-\sqrt{2}D\right]\pa^2\phi_2\pa \phi_2 \right.
\nonumber\\
&&\left.~~~~+\left[(\sqrt{2}-\frac{\a^2}{\sqrt{2}})A
-(\frac{1}{\sqrt{2}} +\sqrt{2} \a
+ \frac{\a
^2}{2\sqrt{2}})C +\frac{\a}{2}D +\sqrt{2}E \right]\pa^3\phi_1
\right.\nonumber\\
&&\left.~~~~+\left[-\sqrt{2}A+
\frac{\a^2}{\sqrt{2}}B +(\frac{1}{\sqrt{2}} +\sqrt{2} \a
+ \frac{\a
^2}{2\sqrt{2}})C +(1+\frac{\a}{2})D -\sqrt{2}F \right]\pa^3\phi_2
\right. \nonumber\\
&&~~~~ \left.-2\sqrt{2}(A-B)(\pa \phi_2)^2\pa \phi_1
\right)e^{\sqrt{2}(\phi_1-\phi_2)} \nonumber\\
{}~~~
\ena
where we have used again various
identities that follow from integration by parts.
We obtain equations
which have a nontrivial solution leading to the quantum-conserved
current
\bea
J^{(3)} &=&(\pa \phi_1)^4+(\pa \phi_2)^4 -6(1+\frac{\b^2}{2\pi})
(\pa \phi_1)^2(\pa \phi_2)^2
+(2+\frac{3\b^2}{2\pi}+\frac{\b^4}{8\pi^2})(\pa^2\phi_2)^2\nonumber\\
&&+6(\sqrt{2}+\frac{3\b^2}{2\sqrt{2}\pi} +\frac{\b^4}{4\sqrt{2}\pi^2})
(\pa \phi_1)^2\pa^2\phi_2 -(4+\frac{6\b^2}{\pi}+\frac{23\b^4}{8\pi^2}
+\frac{3\b^6}{8\pi^3})(\pa^2\phi_1)^2 \nonumber\\
{~}~
\ena

\sect{One-loop charges}

{}From the quantum-conserved currents we can construct corresponding
higher-spin charges. Their existence
has strong implications for the S-matrices of the massive Toda theories
we are considering: the $n$-body S-matrices factorize into products of
2-body elastic S-matrices. Furthermore,
the S-matrix bootstrap principle \cite{2}
implies certain relations between the
charges of the particles participating in the bootstrap \cite{9.5,7.5}.
Conversely,
the violation of these relations, which can be examined once the charges
are given, implies a breakdown of the bootstrap. These are some of the issues
we wish to discuss in the rest of this article. Since we are now interested
in on-shell properties of the theories, we will do our calculations
in terms of massive states and propagators, with the classical masses
and the interactions we shall need obtained from the quadratic and
higher-order parts of the exponentials in the Toda lagrangians.

{}From the current conservation law follows the time-independence of the
corresponding charge operators $Q^{(s)} = \int dx^+J_+^{(s)}$,
hence $Q^{(s)}$ commutes with the Hamiltonian of the system. Single
particle states are therefore eigenstates of the charge operators. Since
the charge $Q^{(s)}$ has helicity $s$,
its action on single-particle states is
\EQ
Q^{(s)} |p      \rangle  = \b^2 \o~  p_+^s |p      \rangle
\EN
(The $\b^2$ factor has been introduced to account for our normalization of
the fields \cite{9}, cf. the $\b^2$ in the propagator, eq. (2.4).)
Acting on a product of wave-packets, the operator $\exp i Q^{(s)}$,
$s>1$, displaces them relative to each other, so that, since $Q^{(s)}$ commutes
with the S-operator, a multiparticle scattering amplitude is equal to one where
well-separated wave-packets scatter pairwise, i.e. the S-matrix factorizes into
products of two-body S-matrices. The elasticity is based on the charge
conservation relation for the process $p_a+p_b + \dots \rightarrow p_f +p_g +
\dots$
\EQ \o_a p_{+a}^s +\o_b p_{+b}^s +\dots =  \o_f p_{+f}^s+ \o_g p_{+g}^s
+\dots
\EN
which can be satisfied, generically, only if the outgoing momenta are
at most a permutation of the incoming momenta.

We will refer to $\o$ as the particle charge and compute it
in terms of on-shell matrix elements $\left\langle p|J^{(s)}(0)|p\right\rangle
\sim p_+^{s+1}$, by
\EQ
\left\langle q|Q^{(s)}|p\right\rangle  =\left\langle
q|\int dx^+J_+^{(s)}(x)|p\right\rangle = 2\pi \d (p_+-q_+)
\left\langle p|J^{(s)}(0)|p\right\rangle
\EN
We are using a normalization
\EQ
\left\langle q|p\right\rangle  =\d (p_1-q_1) = \frac{p_+}{p_0}\d (p_+-q_+)
\EN
We obtain
\EQ
\o =\frac{2\pi p_0}{\b^2 p_+^{s+1}} \left\langle p|J^{(s)}(0)|p\right\rangle
\EN

At the classical level, for the cases discussed in the previous
sections, the charges can be read off from the quadratic parts of the
classical currents. For the $a_2^{(1)}$ theory we observe that
$m_1=m_2$ and the
charge is not diagonal in the $\phi_1$, $\phi_2$ basis. However
$\frac{1}{\sqrt{2}} (\phi_1 \mp i\phi_2)$ are the proper combinations
which diagonalize it and we have, for the spin-2 charge
\EQ
\o^{(0)}_{\pm}=\pm \frac{1}{\sqrt{2}}
\EN
Similarly, in the $a_3^{(2)}$ theory, from the $O(\b^0)$ quadratic part
of the current, we have
\EQ
\o^{(0)}_1=0 ~~~~~,~~~~\o^{(0)}_2=1
\EN
Finally, in the $c_2^{(1)}$ theory,
\EQ
\o^{(0)}_1=-4 ~~~~~,~~~~\o^{(0)}_2=2
\EN

At the quantum level we write $\o =\o^{(0)} + \b^2 \o^{(1)} +\cdots$.
In the following subsections we will compute the one-loop contributions
to these charges. These arise from three sources: one-loop contributions
from the classical parts of the currents,  tree level contributions
from the $O(\b^2)$ parts of the currents, and wave-function renormalization
contributions.

\subsection{The $a_2^{(1)}$ theory}

{}From eq. (3.1) we read the masses and the cubic terms in
the interaction lagrangian
\bea
&&m_1^2=m_2^2=3 \nonumber\\
&&\b^2 {\cal L}^{(3)}= -\frac{\sqrt{2}}{4}\phi_1^3 +\frac{3\sqrt{2}}{4}\phi_1
\phi_2^2
\ena
We compute up to $O(\b^2)$ contributions to $\frac{2\pi
p_0}{\b^2 p_+^3}\left\langle 1|J^{(2)}|2\right\rangle $, with $J^{(2)}$ given
in
 eq. (3.5). They  correspond
to the Feynman diagrams represented in Fig.1 and are given by
\bea
(a)&:& \frac{3\sqrt{2} \b^2}{p_+^3(2\pi)^2}\int d^2k \frac{k_+(p-k)_+p_+}
{[k^2-3][(k-p)^2-3]} \nonumber\\
&&~~~~=-i\frac{\b^2}{2\sqrt{2}\pi}(1-\frac{2\pi}{3\sqrt{3}}) \nonumber\\
(b)&:& \frac{9\sqrt{2}\b^2}{2p_+^3(2\pi)^2}\int d^2k \frac{k_+^3}{
[k^2-3]^2[(p-k)^2-3]} \nonumber\\
&&~~~~=
i\frac{\b^2}{4\sqrt{2}\pi}(\frac{4}{3}-\frac{7\pi}{9\sqrt{3}}) \nonumber\\
(c)&:& i\frac{\sqrt{2}}{2}(1+\frac{\b^2}{4\pi})Z_1^{\frac{1}{2}}Z_2^{\frac
{1}{2}} \nonumber\\
&&~~~~= i\frac{\sqrt{2}}{2}(1+\frac{\b^2}{4\pi})\left[1+\frac{\b^2}{4\pi}(
\frac{\pi}{9\sqrt{3}}-\frac{1}{3})+\dots \right]
\ena
The last contribution contains the $\b^2$ corrections to the
current, as well as one-loop wave-function renormalization factors.
Therefore, up to order $\b^2$ we find the spin 2 charge
\EQ
\o_{\pm} =\pm\left(\frac{1}{\sqrt{2}}+
\frac{\b^2 }{6\sqrt{6}}\right)
\EN
Unlike the nonsimply-laced cases to be discussed presently, quantum
effects do not change the classical charge ratio.

\begin{figure}
\vglue 5cm
\vspace{-1cm}
\caption{Diagrams for the calculation of the charge; the wavy line indicates
insertion of the current.}
\end{figure}

\subsection{The $a_3^{(2)}$ theory}

As in the previous case, we have one-loop contributions from the
classical current, tree-level contributions from the one-loop
corrections to the current, and contributions from wave-function
renormalization. We note that the field $\phi_1$ has vanishing classical
charge.
{}From the lagrangian in eq. (3.8) we read the masses
and the three-point couplings
\bea
&&m_1^2=6 ~~~~,~~~m_2^2=2 \nonumber\\
&&\b^2{\cal L}^{(3)}= -\phi_1^3+\phi_1\phi_2^2
\ena
The relevant part of the current, for computations up to one-loop, is
\EQ
J^{(3)}= 2 \pa \phi_1 \pa \phi_2 \pa^2 \phi_2+
 (1+\frac{23\b^2}{24\pi})(\pa^2\phi_2)^2 -\frac{\b^2}
{24\pi}(\pa^2\phi_1)^2
\EN
For the computation of $\o_1$ we have contributions from the diagrams
in Fig.1 which give
\bea
(a)&:& \frac{4i\b^2}{p_+^4(2\pi)^2}\int d^2k \frac{p_+k_+^2(p-k)_+}{
[k^2-2][(p-k)^2-2]} \nonumber\\
&&~~~~= \frac{\b^2}{12\pi}(1-\frac{4\pi}{3\sqrt{3}}) \nonumber\\
(b)&:& \frac{4i\b^2}{p_+^4(2\pi)^2}\int d^2k \frac{k_+^4}{[k^2-2]^2
[(p-k)^2-2]} \nonumber\\
&&~~~~=-\frac{\b^2}{24\pi}(1-\frac{4\pi}{3\sqrt{3}}) \nonumber\\
(c)&:& -\frac{\b^2}{24\pi}
\ena
Because the classical charge vanishes there is no additional
contribution from wave-function renormalization.
We obtain
\EQ
\o_1 = -\frac{\b^2}{18\sqrt{3}} +O(\b^4)
\EN
For the computation of $\o_2$ we have
contributions from a similar set of diagrams
\bea
(a)&:& \frac{4i\b^2}{p_+^4(2\pi)^2}\int d^2k \frac{(-p_+)k_+(p-k)_+^2}{
[k^2-2][(p-k)^2-6]} \nonumber\\
&&~~~~= \frac{\b^2}{12\pi}(9\ln 3 +\sqrt{3}\pi-15) \nonumber\\
(b)&:& \frac{4i\b^2}{p_+^4(2\pi)^2}\int d^2k \frac{k_+^4}{
[k^2-2]^2[(p-k)^2-6]} \nonumber\\
&&~~~~=- \frac{\b^2}{12\pi}( \frac{9}{2} \ln 3 + \frac{5\sqrt{3}\pi}{6} -\frac
{19}{2})\nonumber\\
(c)&:& (1+\frac{23\b^2}{24\pi})Z_2
 \nonumber\\
&&~~~~= 1+\b^2 (\frac{\sqrt{3}}{36}+\frac{19}{24\pi}) +O(\b^4)
\ena
We obtain
\EQ
\o_2=1+\frac{\b^2}{24\pi}(9\ln 3 +\sqrt{3}\pi +8)+O(\b^4)
\EN
One-loop corrections do not maintain the ratio of the
charges, in particular
the charge of particle $\phi_1$ is no longer zero. As we will
discuss later on, this has significant implications.

\subsection{The  $c_2^{(1)}$ theory}
For this system the masses and relevant couplings
are
\bea
&&m_1^2 =4 ~~~~,~~~m_2^2=8 \nonumber \\
&&\b^2{\cal L}^{(3)} = 2\sqrt{2}\phi_1^2\phi_2
\ena
and the relevant terms of the current are
\EQ
J^{(3)} = 6\sqrt{2}(\pa \phi_1)^2\pa^2\phi_2 -(4+\frac{6\b^2}{\pi})(\pa^2
\phi_1)^2 +(2+\frac{3\b^2}{2\pi})(\pa^2\phi_2)^2
\EN
For the computation of $\o_1$ we have the contributions from Fig.1
\bea
(a)&:& \frac{24i\b^2}{p_+^4\pi^2}\int d^2k \frac{p_+k_+(p-k)_+^2}{[k^2-
4][(p-k)^2-8]} \nonumber\\
&&~~~~=\frac{\b^2}{\pi}(9-3\pi ) \nonumber\\
(b_1)&:& -\frac{32i\b^2}{p_+^4\pi^2}\int d^2k\frac{k_+^4}{[k^2-4]^2[(
p-k)^2-8]} \nonumber\\
&&~~~~=-\frac{\b^2}{\pi}(1+2\ln 2 -\frac{3\pi}{4}) \nonumber\\
(b_2)&:& \frac{16i\b^2}{p_+^4 \pi^2} \int d^2k \frac{(p-k)_+^4}{[k^2-4]
[(p-k)^2-8]^2} \nonumber\\
&&~~~~= -\frac{\b^2}{2\pi}(9-4 \ln 2 -2\pi ) \nonumber\\
(c)&:& -4 -\frac{\b^2}{\pi}[6-(1-\frac{\pi}{4})]
\ena
where the last term consists of a contribution from the
one-loop correction to the current and
an additional contribution from the wave-function renormalization factor.

For the computation of $\o_2$ we have, in a similar manner,
\bea
(a)&:& -\frac{12i \b^2}{p_+^4\pi^2}\int d^2k \frac{p_+^2k_+(p-k)_+}{[k^2
-4][(p-k)^2-4]} \nonumber\\
&&~~~~=-\frac{\b^2}{4\pi}(6-3\pi )\nonumber\\
(b)&:& -\frac{32i\b^2}{p_+^4\pi^2}\int d^2k \frac{k_+^4}{[k^2
-4]^2[(p-k)^2-4]} \nonumber\\
&&~~~~= \frac{\b^2}{8\pi}(8-3\pi ) \nonumber\\
(c)&:& 2+\frac{\b^2}{4\pi}(6-1)
\ena
where, in the very last paranthesis, the $-1$ is a contribution from
wave-function renormalization.

Adding all the contributions we find the one-loop corrections to the charges
\bea
\o_1&=&-4-\frac{3\b^2}
{2\pi}(1+\pi ) +O(\b^4)\nonumber\\
\o_2 &=& 2+\frac{3\b^2}{4\pi}(1+\frac{\pi}{2}) +O(\b^4)
\ena
We note that  here again the quantum corrections do not respect the
classical charge ratio.

\sect{The charge-mass renormalization connection}

It has been established \cite{7.5} that for the simply-laced affine Toda
theories one-loop corrections do not affect the masses  of the particles
except for an overall rescaling. This is not the case in the
nonsimply-laced theories \cite{10}. In this section we discuss the
connection between the quantum corrections to the masses
and the charges in these theories. This connection is intimately related to,
and can be derived from exact S-matrices satisfying
bootstrap relations, but for the
time being we stay at the lagrangian level.

We consider extending the charge
conservation laws in eq. (4.2)
to the three-particle, on-shell vertex function. As we
discuss in the next section, this extension suffers from one notable
failure when applied to the $\left\langle \phi_1 \phi_1 \phi_1\right\rangle $
vertex function in the
$a_3^{(2)}$ theory, due to the presence of an on-shell anomalous
threshold singularity. Postponing discussion of this case, we recall first the
consequences of momentum and spin-3 charge conservation for an on-shell
vertex function $\left\langle \phi_a \phi_b \phi_c\right\rangle $:
\bea
&&p_{+a}+p_{+b}+p_{+c}=0 \nonumber\\
&&\o_a p_{+a}^3 +\o_b p_{+b}^3+ \o_c p_{+c}^3=0
\ena
with $p_+ = \frac{m}{\sqrt{2}}e^{\th}$ in terms of the rapidity.
In particular, for the vertex function
$\left\langle \phi_a \phi_a \phi_b\right\rangle$,
in the frame of reference where the two particles $\phi_a$ have
rapidity $\pm i\th$
while the third particle has rapidity zero (the on-shell
condition forces the momenta to be complex in general) the conservation
laws
read
\bea
2m_a \cos\th &=& m_b \nonumber\\
2\o_a m_a^3 \cos 3\th &=& \o_b m_b^3
\ena
unless the vertex function vanishes.
If $a \neq b$ one derives the following relation:
\EQ
\frac{\o_b}{\o_a}=1-3\frac{m_a^2}{m_b^2}
\EN
(In particular, if $\o_b=0$ either $\o_a=0$ or else the mass ratio must be
$\frac{m_b}{m_a}=\sqrt{3}$.)

We consider now the charge conservation laws in the $c_2^{(1)}$ theory,
applied to the vertex function $\left\langle \phi_1 \phi_1
\phi_2\right\rangle$. At the classical level, with
$\frac{m_1}{m_2}=\frac{1}{\sqrt{2}}$, the above relation implies
$\frac{\o_2}{\o_1}=-\frac{1}{2}$ which is indeed in agreement with the values
of
the classical charges. We extend now these results to the one-loop level. From
Ref. \cite{10} we record the one-loop correction to the mass ratio
\EQ
\frac{m_1^2}{m_2^2}= \frac{1}{2} -\frac{\b^2 }{32}
\EN
which, when substituted into eq. (5.3) gives
\EQ
\frac{\o_2}{\o_1} = -\frac{1}{2} +\frac{3\b^2 }{32}
\EN
which is indeed in agreement with the one-loop corrections to the
charges that we have computed in eq. (4.22).

The same procedure can be used for the vertex function
$\left\langle \phi_2 \phi_2 \phi_1\right\rangle $ of the $a_3^{(2)}$ theory.
At the classical level, using the mass ratio $\frac{m_2}{m_1} =
\frac{1}{\sqrt{3}}$
we conclude that either the coupling $\phi_1\phi_2^2$ is zero, or the
charge $\o_1 =0$. In fact the coupling is not zero, but the classical
charge vanishes. We consider now the situation at the one-loop level.
Here, from Ref. \cite{10} we have,
\EQ
\frac{m_2^2}{m_1^2}= \frac{1}{3} +\frac{\b^2}{54\sqrt{3}}
\EN
Using eq. (5.3) we obtain
\EQ
\frac{\o_1}{\o_2} = -\frac{\b^2}{18\sqrt{3}}
\EN
which is again in agreement with the one-loop corrections in eqs. (4.15),
(4.17).

We consider now the vertex function for
three identical particles. Setting  $a=b$
in eq. (5.2), the first equation gives $\th =\frac{\pi}{3}$ and
when substituted in
the second equation it implies $\o_a=0$.
Therefore, either the three-point function of a field $\phi$ is zero on shell,
or the corresponding spin-3 charge vanishes.  More generally this is the case
for any charge unless its spin is $s=6n\pm 1$ for some integer $n$.
As an example, we note that
in the $c_2^{(1)}$ theory, where the spin-3 charge of the fields is not
zero, the couplings $\phi_1^3$ and $\phi_2^3$ are indeed zero.
We have checked that this also holds at the one-loop level.

On the other hand, in the $a_3^{(2)}$ theory, following the same reasoning
we would conclude that either the on-shell vertex function
$\left\langle \phi_1 \phi_1 \phi_1\right\rangle $ vanishes identically, or the
corresponding charge $\o_1=0$. In fact, at the classical level the
coupling $\phi_1^3$ is present but the classical charge vanishes so
that there is no contradiction.
However at the quantum level the charge $\o_1$  is not zero,
and therefore the conservation arguments that lead to eq. (5.2) must break
down.
We examine the situation in the next section.

\sect{The breakdown of charge conservation}

Given a spin $s$ conserved current, the corresponding on-shell charge
conservation laws can be derived by the following argument \cite{11}: using
$\pa_-J^{(s)}_++\pa_+J^{(s)}_-=0$ we have
\EQ
-\int d^2x e^{i\e_-x^-}\pa_-J_+^{(s)}(x) =\int
d^2xe^{i\e_-x^-}\pa_+J_-^{(s)}(x)
=\int d^2x \pa_+\left( e^{i\e_-x^-}J_-^{(s)}(x) \right) =0
\EN
The relation
\EQ
\left\langle out| \int d^2x e^{i\e_-x^-}\pa_-J_+^{(s)}(x)|in\right\rangle  =0
\EN
has the graphical interpretation shown in Fig.2: one inserts the
operator $\pa_- J_+^{(s)}$, carrying momentum $\e=(0,\e_-)$
 in all possible ways in
the Feynman diagram for the process $\left\langle out|in\right\rangle $ and the
result must be zero.
\begin{figure}
\vglue 5cm
\caption{Illustrating the derivation of charge conservation}
\end{figure}

When the operator is
inserted in an external line with on-shell momentum $p$
one gets a factor
proportional to $\o p_+^{s+1}$ from the current,
a  factor $\e_-$ from the derivative, and a factor with a pole singularity
as $\e_- \rightarrow 0$
\EQ
\frac{1}{(p+\e)^2-m^2}=\frac{1}{2p_+\e_-}
\EN
from the additional propagator with on-shell external momentum.
When the current is inserted into an internal line, one normally gets
contributions which do not have a pole singularity in the limit $\e_-
\rightarrow 0$ but still have the numerator factor $\e_-$. Therefore in
the limit $\e_-
\rightarrow 0$ one finds
$(\sum_i \o_i p_{i+}^s)\left\langle out|in\right\rangle =0$ implying that
either the sum of the charges is zero
or the corresponding on-shell amplitude vanishes.

A priori this result holds also for the three-point function,
although momentum conservation and the on-shell conditions require that
the momenta be complex. In particular, in two dimensions,
it is implied by the S-matrix bootstrap.
However, precisely in two dimensions, and for example in the
case of the $\left\langle \phi_1 \phi_1 \phi_1\right\rangle $ on-shell vertex
function of the
$a_3^{(2)}$ theory, the argument breaks
down; contributions which come when the
current is inserted inside the Feynman diagram have pole singularities
as $\e_- \rightarrow 0$ because for the particular values of the masses the
diagrams represented in Fig.3b,c  have anomalous threshold singularities
precisely when the three external lines are on the mass-shell, as follows from
standard dual diagram analysis \cite{7.5,9}. Thus, as already suggested by the
contradiction we found in the previous section, the naive charge conservation
breaks down, because in the limit $\e_- \rightarrow 0$ there are additional
contributions to the (integrated) current conservation law.
In this section we demonstrate explicitly this phenomenon for the
particular example of the $\left\langle \phi_1 \phi_1 \phi_1\right\rangle $
vertex.
\begin{figure}
\vglue 5cm
\vspace{-1cm}
\caption{Diagrams showing the breakdown of the naive charge conservation law}
\end{figure}

At the one-loop level we need consider the set of diagrams illustrated
in Fig.3, carry out the computation for $\e_- \neq 0$, and then take the
limit $\e_- \rightarrow 0$. When the current is
inserted in the external $\phi_1$ lines the
classical part gives zero, since $\o_1^{(0)} =0$ and we are left with a
contribution proportional to
$
\o_1^{(1)}( p_{+}^3+q_+^3+k_+^3)
$
times the classical vertex function.
When the current is inserted in an
internal $\phi_2$ line, or at a vertex containing the $\phi_2$ field,
we get a contribution from the classical part.
Thus, we must compute the one-loop diagrams in Fig.3b,c
representing
\EQ
\langle p,q,k| \int d^2x e^{i \e_-x^-}\pa_-[2 \pa \phi_1 \pa \phi_2 \pa^2\phi_2
+(\pa^2\phi_2)^2 ]|0\rangle
\EN
When $\e_- \neq 0$ these diagrams are not at their anomalous
threshold value.
However, in the limit $\e_- \rightarrow 0$ the anomalous threshold
singularities become  simple poles in $\e_-$ and this cancels the corresponding
$\e_-$ numerator factor leading to a finite contribution to the charge
conservation law.

We evaluate the Feynman integrals corresponding to the diagrams in
Fig.3, with the external lines on-shell, and
with the current insertion introducing a momentum $\e = (0,\e_-)$. Each
diagram corresponds to three possible places for the current insertion.
We use
\bea
&&2p_+p_- =2q_+q_-=2k_+k_- =m_1^2=6 \nonumber\\
&&p_++q_++k_+=0   ~~~~,~~~p_-+q_-+k_- =\e_- \nonumber\\
&&p_+^3+q_+^3+k_+^3 =\frac{3 m_1^3}{2\sqrt{2}} + O(\e_-)
\ena
where for the last relation we chose the rest frame of one of the particles.

The insertion of the current into the external lines leads to a
contribution
\EQ
(a)=\frac{12 \e_-p_+^4 \o_1 }{(p-\e)^2-m_1^2} +{\rm cyclic}~(p,q,k)
=-6(p_+^3+q_+^3+k_+^3)\o_1
\EN
If this were the only contribution, from eq. (6.2) we would conclude that
$\o_1$ must vanish. However,
$\o_1= -\frac{\b^2}{
18\sqrt{3}}$, cf. eq. (4.15), and using eq. (6.5) we obtain
\EQ
(a)=3\b^2
\EN

The diagram in Fig.3b leads to the integral
\EQ
(b)=\frac{8i\b^2}{(2\pi)^2}\int d^2l
 \frac{\e_-q_+^2(l+q)_+l_+}{[(l-k)^2-2][(l+q-\e )^2-2]
[l^2-2]} +{\rm cyclic}~(p,q,k)
\EN
whereas the diagram in Fig.3c gives
\EQ
(c)=-\frac{16\b^2}{(2\pi)^2}\int d^2l
 \frac{\e_-l_+^4}{[l^2-2][(l+q)^2-2][(l+p+q)^2-2]
[(l+\e )^2-2]} +{\rm cyclic}~~(p,q,k)
\EN
The numerator factors correspond to the derivatives in the current.
The integrals can be computed using, for example, the partial
fractions procedure described in Ref. \cite{12} which reduces them
to self-energy integrals. After a not-insignificant amount of
algebra we find, in the limit $\e_- \rightarrow 0$
\EQ
(b)= -\frac{3\b^2 }{2\sqrt{6}}m_1^3 = -9\b^2
\EN
and
\EQ
(c)= \frac{\b^2}{\sqrt{6}}m_1^3 = 6\b^2
\EN
The sum gives $(a)+(b)+(c)=0$, which is the correct consequence of eq. (6.2).
Thus, the presence of the anomalous threshold singularities modifies
the naive charge conservation law which would have led to the
requirement $\o_1=0$ or $\left\langle \phi_1 \phi_1 \phi_1\right\rangle =0$.

We may expect a similar breakdown of charge conservation to occur in other
theories provided two ingredients are present: masses and couplings such
that an on-shell vertex function has an anomalous threshold singularity,
and terms in the current such that diagrams as in Fig.3b,c exist and give
a nonzero contribution. In particular, this must be the case in many
nonsimply-laced theories where, as discussed in Ref. \cite{8}, the
bootstrap does not hold at all S-matrix poles. A similar problem seems
to be absent in the simply-laced theories.

\sect{Conclusions}

In this paper we have studied the quantum properties of higher-spin
currents of the affine Toda theories $a_2^{(1)}$, $a_3^{(2)}$ and $c_2^{(1)}$.
These theories are
classically integrable, i.e. currents exist which are conserved by virtue of
the classical field equations. We have studied the spin-2 current for
the $a_2^{(1)}$ and the spin-3 current for the $a_3^{(2)}$ and $c_2^{(1)}$
theories.
We have shown that the conservation laws
are maintained at the quantum level, although the currents have to be
modified by
higher-order corrections.

The existence of quantum-conserved currents for the affine Toda theories
implies the existence of factorizable, elastic S-matrices. For the
simply-laced cases exact S-matrices
were proposed some time ago and checked to agree with those computed
from the Toda lagrangians to some low orders of perturbation theory
\cite{7,7.5}.
For the nonsimply-laced cases, e.g. the $a_3^{(2)}$ and the $c_2^{(1)}$
theories, the existence of quantum-conserved currents was somewhat
unexpected. Quantum corrections distort the classical mass spectrum
in a manner that seemed incompatible with the existence of exact,
factorizable S-matrices, and indeed attempts to construct them based
on the usual bootstrap procedures had failed. However, stimulated by
the existence of conserved currents, we were able to determine
the S-matrices for the nonsimply-laced theories, by using
more general constructions than hitherto attempted \cite{8}. In particular,
the realization that anomalous thresholds may modify the conservation
relations for vertex functions made it clear that there are situations
when the bootstrap principle has to be relaxed.

For situations where the
S-matrix bootstrap is valid in the form
\EQ
S_{cd}(\th ) = S_{ad}( \th + \bar{\th}^b_{ac}) S_{bd}(\th - \bar{\th}^a_
{bc})
\EN
(here the S-matrix element $S_{ab}(\th )$ has a simple pole corresponding to
particle $c$)
the charge conservation follows by taking the logarithm of the above
relation and inserting the Fourier expansion \cite{7.5}
\EQ
\varphi_{ab}(\th ) = i\frac{d}{d\th } \ln S_{ab}(\th ) = \sum _{s=0}
^{\infty} e^{-s|\th |}\varphi _{ab}^{(s)}
\EN
so that eq. (7.1) gives
\EQ
\varphi_{cd}^{(s)} = \varphi_{ad}^{(s)}e^{-s\bar{\th}^b_{ac}}
+\varphi_{bd}^{(s)}e^{s\bar{\th}^a_{bc}}
\EN
In general the
matrices $\varphi_{cd}^{(s)}$ have rank $1$ and are related to
the charges by
\EQ
\frac{\g_a^{(s)}}{\g_b^{(s)}} = \frac{\varphi_{ac}^{(s)}}{\varphi_{bc}^{(s)}}
\EN
where $\g_a^{(s)} = \o_ap_{a+}^s$. Eq. (7.3) expresses the charge conservation
law in a three-particle vertex function.

Failure of the conservation law in eq. (7.3), as in the example
of Section 6, implies corresponding failure of the
bootstrap. Indeed, in constructing the exact S-matrices in Ref. \cite{8},
 we found that
the bootstrap should not be applied at certain simple poles of the
S-matrix because these poles, unlike the usual simple particle
poles, are due to an interplay between the usual poles and anomalous
threshold singularities. Obviously, such singularities modify the
bootstrap conditions in the same manner they modify  the charge
conservation laws as discussed in Section 6.
However, although eqs. (7.1) and (7.3) may not be valid in every channel
and for all vertex functions, there are usually enough valid bootstrap
conditions so that eq. (7.4) does hold
and expresses particle charges in terms of the exact S-matrix.

In Ref. \cite{8} we have shown, using the
explicit expressions for the S-matrices
 that for the $a_{2n-1}^{(2)}$
theories the charges obtained from eq. (7.4)
satisfy (using also $p_+ =\frac{m}{\sqrt{2}}e^{\th}$)
\EQ
\frac{\g_a^{(s)}}{\g_n^{(s)}} = \frac{m_a^s \o_a^{(s)}}{m_n^s \o_n^{(s)}}
=(-1)^{\frac{s-1}{2}} 2 \sin
\frac{\pi s a}{H}
\EN
with $H = 2n-1 + \frac{\b^2}{4\pi} +O(\b^4)$
while for the $c_n^{(1)}$ theories they satisfy
\EQ
\frac{\g_a^{(s)}}{\g_{n}^{(s)}} = \frac{m_a^s \o_a^{(s)}}{m_n^s \o_n^{(s)}}
=\frac{ \sin
\frac{\pi s a}{H}}{ \sin \frac{\pi s n}{H}}
\EN
with $H = 2n + \frac{\b^2}{2\pi} +O(\b^4)$.
For our specific examples, using also the one-loop corrected masses in
eqs. (5.6) and (5.4), this leads to
\bea
\frac{\o_1}{\o_2} &=& -\frac{\b^2}{18\sqrt{3}}\nonumber\\
\frac{\o_1}{\o_2} &=& -2(1+\frac{3\b^2}{16})
\ena
for the $a_3^{(2)}$ and the $c_2^{(1)}$ theories respectively, which is
indeed consistent with our explicit calculations.

We conclude with the following remark: the construction of quantum-conserved
currents and charge operators insures the existence of factorized,
elastic S-matrices for the theories we have considered. Aside from
algebraic complexity, we believe the same construction holds for all
simply-laced and nonsimply-laced Toda theories. In Refs. \cite{7,7.5} for
the simply-laced cases, and in Ref. \cite{8} for the nonsimply-laced
cases, exact S-matrices have been determined and perturbative checks
have been performed to identify them with those for specific Toda
theories. To make the identification precise, it is necessary to
specify the {\em quantum} form of the Toda lagrangians. In our
construction of the conserved currents we have assumed that the
exponentials are normal-ordered, and the Ka$\check{c}$ labels
are chosen so that all one-point functions vanish.
This last restriction is not crucial:
it plays no role in checking current conservation, and can always be
achieved by shifting the fields.
However,
it would seem that any other modification
of the lagrangians, e.g. the addition by hand of finite mass counterterms,
would spoil the current conservation and bring into question the
existence of an exact S-matrix for the corresponding theory.

\noindent {\bf Acknowledgments.} We thank P. Dorey and R. Sasaki for
discussions.

\end{document}